# Planar Waveguide Hybrids for Very High Power RF[*]

C.D. Nantista, W.R. Fowkes, N.M. Kroll[#], and S.G. Tantawi[‡]

Stanford Linear Accelerator Center, P.O. Box 4349, Stanford, CA 94309

## Abstract

Two basic designs have been developed for waveguide hybrids, or 3-dB couplers, capable of handling hundreds of megawatts at X-band. Coupling is provided by one or two connecting waveguides with h-plane junctions and matching elements. In the former case, the connecting waveguide supports two modes. Small apertures and field-enhancing e-bends are avoided to reduce the risk of rf breakdown. The h-plane symmetry also allows the use of over-moded rectangular waveguide in which the height has been increased to reduce field amplitudes without affecting the scattering matrix. The theory and designs are presented, along with the results of prototype tests of functionality and power-handling capability. Such a device is integral to the rf pulse compression or power distribution system [2] of the Next Linear Collider (NLC) [1] for combining, splitting, and directing power. This work was motivated by the observation of rf breakdown at power levels above 200 MW in conventional and modified magic-T's.

*presented at the 1999 IEEE Particle Accelerator Conference*

*New York, New York, March 29-April 2, 1999*

---

[*] Work supported by Department of Energy contract DE–AC03–76SF00515

and grant DE-FG03-93ER40695.

[#] Also University of California, San Diego, La Jolla, CA 92093.

[‡] Also with the Communications and Electronics Department, Cairo University, Giza, Egypt.

# I. INTRODUCTION

The design of the Next Linear Collider (NLC) [1] includes plans for powering the high-gradient accelerator structures of the main linacs with 11.424 GHz X-band klystrons through a pulse compression or power distribution system [2]. In such a system pulsed rf will need to be combined, split, or directed at peak power levels reaching 600 MW. A basic component required is a waveguide hybrid, or 3-dB directional coupler, capable of handling very high power levels. Prototype rf systems have employed conventional, matched magic T's in WR90 (0.9"x0.4") waveguide. As power levels were increased, these proved inadequate and a modified design was developed, in which the matching post in the waveguide junction was replaced with a thick fin. While simulations showed this design to have lower field strengths, it's reliability proved to still be inadequate. These magic T's exhibited frequent rf breakdown at power levels above 200 MW, which inspection showed to occur primarily at the mouth of the e-bend [3].

With this motivation, we have subsequently developed two novel planar hybrid designs capable of reliably handling hundreds of megawatts of peak power at X-band (11.424 GHz). These each consist of four rectangular waveguide ports, which operate in the $TE_{10}$ mode, connected through four or two h-plane T-junctions, yielding, respectively, a two-rung ladder or an "H" geometry. In the latter case, the single connecting waveguide carries two modes. Figure 1 illustrates the two design geometries.

Small apertures, slots, and field-enhancing e-bends are avoided to reduce the risk of rf breakdown. Matching features maintain the translational symmetry of these cross-sections. Electric fields terminate only on the flat top and bottom surfaces. This h-plane symmetry also allows the use of over-moded rectangular waveguide in which the height has been increased to reduce field amplitudes without affecting the scattering matrix. Both are quadrature hybrids (i.e. the coupled port fields are 90° out of phase), and directly opposite port pairs are isolated. That is, the scattering matrices, with properly chosen, symmetric reference planes and the indicated port numbering, are of the form

$$\mathbf{S} = \frac{1}{\sqrt{2}} \begin{bmatrix} 0 & 1 & -i & 0 \\ 1 & 0 & 0 & -i \\ -i & 0 & 0 & 1 \\ 0 & -i & 1 & 0 \end{bmatrix}.$$

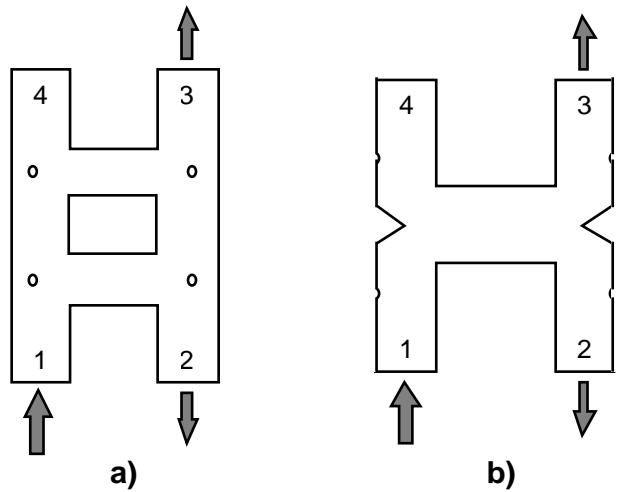

**Figure 1.** Schematic of the h-planar geometries of the a) two-rung ladder and b) "H" hybrid designs. Power-flow arrows indicate output ports for the indicated input port.

## II. TWO-RUNG LADDER HYBRID

The two-rung ladder hybrid is basically a realization in rectangular waveguide of R.H. Dicke's coaxial-line synthesis of a biplanar magic T [4], or of a modified microstrip branch-line hybrid. In transmission line theory, this device requires the distances between all adjacent junctions to be (n/2 plus) one quarter wavelength and the two connecting lines to have a characteristic admittance that is $\sqrt{2}$ times that of the main lines. The resulting circuit can be shown to yield the desired scattering matrix.

In waveguide, whose width is not small compared to a wavelength, the two-dimensional geometry necessitates matching elements in the T-junctions to adjust the complex impedances. A mode matching code was used to determine the radius and placement of a post that would yield a three-port junction with the desired scattering matrix, from which the hybrid ring circuit could then be constructed. The fields were expanded in cylindrical harmonics about the post in the junction region and in normal modes in the rectangular port regions. The full hybrid design was verified with the finite-element field solver HFSS. The peak field amplitude was found to be 44.1 MV/m at 300 MW with a waveguide height of 0.900 inch.

## III. "H" HYBRID

The "H" hybrid can be viewed as a variation of the above with the two connecting waveguides collapsed into one in which two modes are utilized. Thus, in a transmission line picture, the requirement of two connecting lines is not circumvented. The coupling mechanism is actually the same as that of the Riblet short-slot coupler [5], although this geometry provides separated ports and no sharp-edged wall interruptions.

The connecting guide is wide enough that both the $TE_{10}$ and the $TE_{20}$ modes can propagate, and these are excited with comparable amplitudes by the fields of a single port. They are excited with a relative phase such that their fields add constructively on the side nearest the input port and destructively on the other side. If they were to slip in phase by $\pi$ radians, the $TE_{10}$ wave would enhance the opposite lobe of the $TE_{20}$ wave, sending the power out the farthest port. To get a 3-dB split, therefore, the total phase lengths for these two modes through the connecting guide must differ by an odd multiple of $\pi/2$.

The T-junctions in this design have been matched by shaping the walls with blunt, triangular protrusions at the symmetry plane, rather than with free-standing posts. The result is essentially a side-wall coupler with the common wall removed and two back-to-back mitred 90° bends at either end. The connecting guide must be narrower than twice the standard guide width in order to keep the $TE_{30}$ mode cut off. Simple mitred bends used in a preliminary design therefore led to narrow ports (half the connecting guide width). To accommodate standard-width ports and avoid the added length of width tapers, a small vertical ridge was placed in each port to match into an effective asymmetric mitred bend. The $TE_{20}$ mode is thus matched independently at each junction. The width and length of the connecting guide are adjusted to simultaneously meet the above phase length difference requirement and cause the small $TE_{10}$ mismatch at the two junctions to cancel. HFSS was used extensively in the design process to calculate scattering matrices. The peak field of the final design was found from simulation to be 39.5 MV/m at 300 MW with a waveguide height of 0.900 inch.

## IV. TESTS

A copper high-power vacuum-flanged prototype of each of these hybrids has been built and tested. They were both made over-moded, the two-rung ladder in 0.9 inch square guide and the "H" in double-height (0.9"x0.8") guide. This necessitated height tapers at the ports for compatibility with our test setup and other WR90-based components. The top and bottom were tapered symmetrically with one-inch long half-cosine tapers. An HP 8510C Network Analyzer was used to measure the scattering matrix parameters in the vicinity of the design frequency. The results over a 500 MHz span are presented in Figure 2.

The measured insertion losses, when corrected for the predicted loss of the flange adaptors used and WR90 curved h-bends built onto two ports of the "H" hybrid to accomodate a particular installation, give ~1.5% for the two–rung ladder hybrid and ~0.9% for the "H" hybrid. We define this loss as $1-(|S_{21}|^2+|S_{31}|^2)$. It is dominated by ohmic loss. For the two-rung ladder, the reflected signal and isolation at 11.424 GHz were both about -26 dB, accounting together for 0.49% misdirected power. For the "H" they were -33 dB and -37 dB, respectively, accounting for 0.07% misdirected power. The measured coupling at 11.424 GHz, corrected for loss (i.e. $10\log[|S_{31}|^2/(|S_{21}|^2+|S_{31}|^2)]$), was -3.19 dB for the former hybrid and -2.96 dB for the latter (the ideal being -3.01 dB).

Finally, Figure 2 shows the "H" hybrid to have a significantly broader bandwidth, as one would expect from the more compact geometry.

The hybrids were later high-power tested in the pulse compression system of the Accelerator Structure Test Area (ASTA) [3], where they were processed with pulsed rf to peak power levels exceeding 400 MW in 150 ns pulses and performed successfully without breakdown problems or excessive X-ray production.

## V. CONCLUSIONS

In response to the problem of rf breakdown in multi-hundred-megawatt X-band rf systems being developed for a next generation linear collider, we have conceived and produced two new types of rectangular waveguide hybrid, with relatively open interiors and completely two-dimensional designs, perpendicular to the electric field lines. The latter feature makes their circuit properties independent of height, allowing for their construction in over-height waveguide to reduce fields. Prototypes of both designs performed similarly and quite well. The "H"-shaped hybrid has the advantage of broader bandwidth and is more compact. HFSS simulation suggests that it has peak fields lower by about 11% for a given power flow and waveguide height. The absence of free-standing matching elements may also be an advantage with regard to cooling.

One goal in our component development program is to limit surface fields at anticipated power levels to values below 40 MV/m in order to avoid rf breakdown problems. For a power flow of 300 MW in one port, both hybrid designs meet or approach this goal in square guide (0.9"x0.9"). By contrast, our original and modified magic T's had peak fields of approximately 80 MV/m and 63 MV/m, respectively, at this power level and could not be made over-moded in height.

For testing and for their intended use, we required standard, single moded ports on our prototypes, for which the peak field at 300 MW is 49 MV/m. Smooth height tapers were incorporated at the ports to bring the peak fields in the interior of these devices, where standing waves cause some enhancement, below the peak port field. With reference planes taken just inside these tapers, the hybrids proper are thus over-moded. To take full advantage of these hybrid designs, one would not normally use single-moded ports, but remain over-moded, perhaps matching into a $TE_{11}$ mode in circular waveguide.

To comfortably handle 600 MW, the hybrid height would have to be increased to 1.75 inches. In such waveguide, taper design becomes non-trivial because the $TE_{12}$ mode can propagate. Mode conversion due to mechanical imperfections also becomes more of a concern as a device becomes more over-moded. It may therefore be preferable to use a configuration which incorporates two hybrids to further increase power-hancdling capacity.

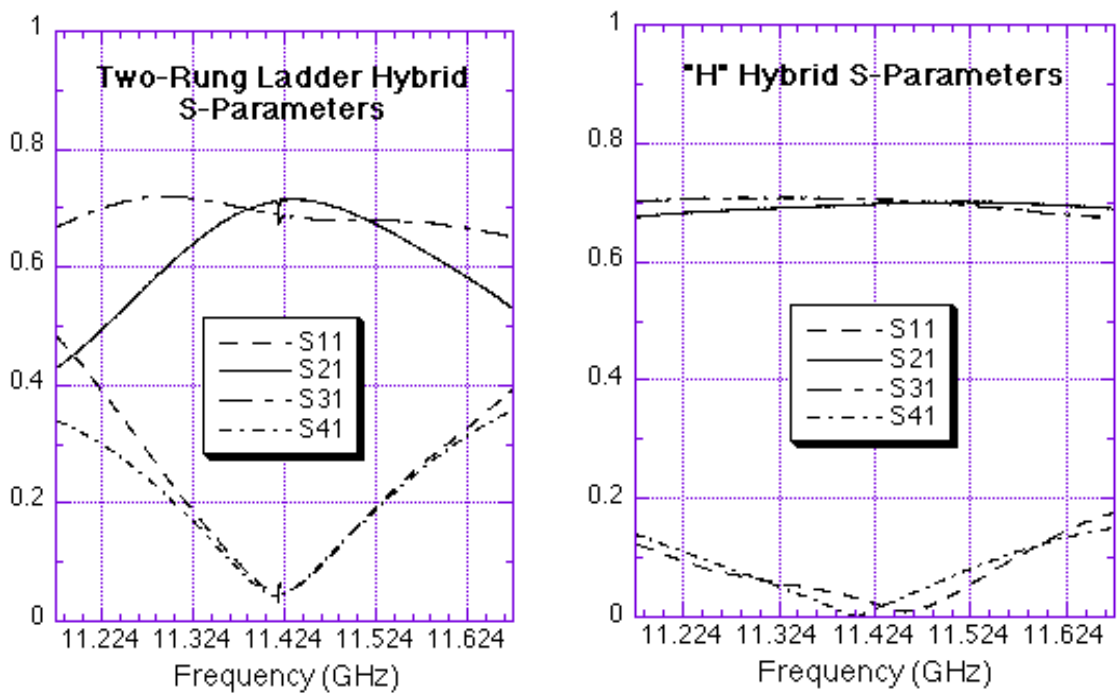

**Figure 2.** Scattering matrix elements for our two hybrid prototypes measured over a frequency range of 500 MHz centered on the design frequency of 11.424 GHz.